# Nonequilibrium Casimir–Lifshitz Friction force and Anomalous Radiation Heating of a Small Particle

G. V. Dedkov

Kabardino-Balkarian State University, Nalchik, 360004, Russia

This paper presents the results of calculating Casimir-Lifshitz friction force and heating rate of a small metal particle moving above a metal surface (thick plate) in the case of their different local temperatures. The case of normal nonmagnetic metals (Au) is considered. There is a strong interplay of temperatures, particle velocity and separation distance resulting in an anomalous direction of the heat flux between the bodies and a peak temperature dependence of the friction force at sufficiently low temperatures. In particular, a "hot" moving particle can additionally receive heat from a "cold" surface. The conditions for experimental measurement of these effects are discussed.

The electrical in origin Casimir–Lifshitz forces arise between neutral condensed dissipative bodies without permanent polarizations due to spatial correlations of quantum and thermal fluctuations of polarization and magnetization of bodies and a vacuum, or, equivalently, due to the fluctuations of the electromagnetic waves between neutral bodies [1–3]. Nonequilibrium effects in thermal radiation is another area of manifestation of fluctuating electromagnetic fields [5, 6, 7–9]. The theory of electromagnetic fluctuations developed by Rytov [5, 6] provided a powerful tool to deal with thermal radiation and dispersion forces in different fields of science [7]. During several decades, a large body of literature has been devoted to the low-temperature behaviour and entropy anomaly of static Casimir-Lifshitz force between metallic and dielectric bodies (see a recent review [4]). Much attention has also been drawn to the large enhancement of heat transfer between closely spaced bodies in the near field. The extensive reviews on this issue see in [11, 12–15].

Dynamic or combined dynamic-thermal disequilibrium may cause new interesting phenomena, such as temperature-dependent Casimir-Lifshitz (CL) friction and quantum friction between bodies in relative motion [16–20], radiation in a vacuum [21–27], etc. [28–34]. In [35, 36], we pointed to the fascinating low-temperature dependence of the CL friction force caused by the strong temperature dependence of the damping frequency for normal nonmagnetic metals. Though at room temperatures the friction force is very small [13, 25, 38], new perspectives in measurements of both conservative and dissipative CL forces are opening up due to recent low-temperature experiments [39, 40]. So far, dynamic effects of radiative heat transfer have not been discussed, although the general expressions for the rate of heat exchange in dynamic configurations include the velocity factors [13, 25, 41]. In particular, a situation may arise when a "hot" moving particle receives heat from a "cold" plate. The second law of thermodynamics is not violated in this case, since the system including the bodies in relative motion and fluctuational electromagnetic field is nonequilibrium.

This paper presents the results of numerical calculations of the CL friction force and radiation heat transfer (RHT) for a spherical particle moving above a thick plate (both from normal nonmagnetic metals), depending on the temperatures, separation distance, and particle velocity. The choice of materials (metals like gold) is due to the fact that the anomalous direction of RHT manifests itself more noticeably at sufficiently low temperatures and low speeds, which favours experiments. In addition, metals have a significant temperature dependence of the dielectric properties in the low-frequency spectral range, which is responsible for the RHT due to an increase in the density of electromagnetic modes. In contrast, in dielectric materials like $SiO_2$ or SiC, the absorption of electromagnetic radiation is "tied" to the peaks of the imaginary part of the permittivity, which are in the infrared range of the spectrum. Therefore, to implement anomalous RHT between dielectric bodies, much higher velocities and rigid restrictions on the relationship between various parameters are required, which makes its experimental observation difficult.

The configuration is shown in Fig. 1. We will use the permittivity $\varepsilon(\omega)$ of both materials in the form of the Drude expression

$$\varepsilon(\omega) = 1 - \frac{\omega_p^2}{\omega(\omega + i\nu)}, \qquad (1)$$

where $\omega_p$ is the plasma frequency and $\nu$ is the damping parameter. We also assume that $\nu$ depends on the temperature $T$ via the Bloch–Grüneisen law [42] ($\theta = 175$ K and $\omega_p = 9.03$ eV for gold)

$$\nu_{BG}(T) = 0.0212(T/\theta)^5 \int_0^{\theta/T} x^5 \text{sh}(x/2)^{-2} dx, \text{(eV)}. \qquad (2)$$

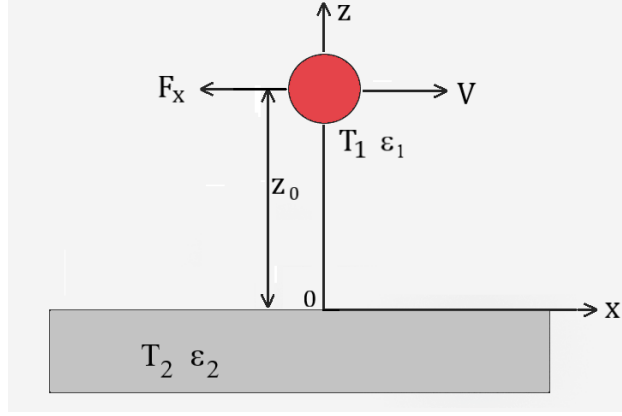

Fig. 1. Particle-plate configuration.

As noted by several authors [13, 25, 35, 43], in a separation range of $1 \div 100$ nm, the determining role in the CL friction force and the rate of heat transfer between metallic bodies is played by terms that include the products of the reflection coefficients $\Delta_m$ corresponding to S-polarized electromagnetic waves and (or) magnetic polarizabilities $\alpha_m$:

$$\Delta_m = \frac{q_0 - q}{q_0 + q}, \quad \alpha_m = \frac{2\pi R^3}{15}\left(\frac{\omega R}{c}\right)^2 (\varepsilon(\omega) - 1), \tag{3}$$

where $q_0 = \sqrt{k^2 - \omega^2/c^2}$, $q = \sqrt{k^2 - \varepsilon(\omega)\omega^2/c^2}$, $\mathbf{k} = (k_x, k_y)$ is the two-dimensional wave vector in the plane of the surface, $R$ is the particle radius. The above expression for $\alpha_m$ is valid in the limit $R \ll \delta$ (with $\delta$ being the skin depth).

General relativistic expressions for the dissipative CL force experienced by a dipole particle were obtained in [41, 44]. Using the notation from Ref. [25] (Eq. (145)), in the limit $V \ll c$ with retardation effects accounted for and retaining only the contribution due to the magnetic polarization of the particle, we rewrite this formula in the form (see Supplemental material)

$$F_x = \frac{\hbar}{\pi^2}\int_0^\infty d\omega \int d^2k\, k_x A_m \alpha_m''(\omega_+)\mathrm{Im}\left(\Delta_m(\omega)\frac{e^{-2q_0 z_0}}{q_0}\right)\left[\coth\left(\frac{\hbar\omega_+}{2T_1}\right) - \coth\left(\frac{\hbar\omega}{2T_2}\right)\right], \tag{4}$$

where $A_m = k^2 - \frac{\omega^2}{c^2} + \frac{\omega_+^2}{2c^2}$, $\omega_+ = \omega + k_x V$, $z_0$ is the distance between the center of the particle and the surface of the plate, $\alpha_m''$ is the imaginary part of $\alpha_m$, temperatures $T_1$, $T_2$ and indices 1,2 of the quantities introduced below (such as $v_i = v_i(T_i)$ in (1), etc.) correspond to a moving particle and a resting plate, respectively.

Under the same conditions, the expression for the rate of particle heating (see Eq. (147) in [25]) takes the form

$$\frac{dQ}{dt} = -\frac{\hbar}{\pi^2}\int_0^\infty d\omega \int d^2k \,\omega_+ A_m \alpha_m''(\omega_+) \mathrm{Im}\left(\Delta_m(\omega)\frac{e^{-2q_0 z_0}}{q_0}\right)\left[\coth\left(\frac{\hbar\omega_+}{2T_1}\right) - \coth\left(\frac{\hbar\omega}{2T_2}\right)\right] \quad (5)$$

One can see that Eq. (5) reduces to (4) when replacing $\omega_+$ for $-k_x$. At this stage, Eqs. (4), (5) describe the contributions from both traveling and near field electromagnetic waves.

The effect of the anomalous RHT is already seen from (5). Let $T_1, T_2 \to 0$, but $T_1 > T_2$, then Eq. (5) reduces to [25]

$$\frac{dQ}{dt} \cong \frac{4}{\pi^2}\int_0^\infty dk_x \int_0^\infty d k_y k e^{-2kz_0} \int_0^{k_x V} d\omega \, \Delta_m''(\omega)(\omega - k_x V)\alpha_m''(\omega - k_x V) > 0, \quad (6)$$

since $\alpha_m''(\omega)$ is an odd function. The anomalous RHT is due to photons with $k_x < 0$, when $\omega_+ = \omega + k_x V < 0$ at $\omega < |k_x|V$. With decreasing temperature, the characteristic frequencies of absorption of thermal radiation also decrease, which contributes to the fulfillment of the above condition at lower velocities. Another positive factor is associated with an increase in the density of low-frequency electromagnetic modes.

At $z_0 < 100$ nm, traveling modes make an insignificant contribution to Eqs. (4), (5) [35]. Therefore, in numerical computations, we can take into account only the range $\omega < kc$, introducing new variables $\omega = v_1 t$, $k = (\omega_p/c)\sqrt{y^2 + \beta_1^2 t^2}$, with $\beta_1 = v_1/\omega_p$ and parameters $\alpha_i = \hbar v_i/T_i$ ($i = 1,2$), $\gamma = v_1/v_2$, $\lambda = \omega_p z_0/c$, $w = T_1/T_2$. Moreover, it is convenient to use the polar coordinates $(k, \phi)$ in the plane $(k_x, k_y)$, denoting $\omega_+ = v_1 t_+$ with $t_+ = t + kV\cos\phi/v_1$. Then Eqs. (4), (5) take the form

$$F_x = \frac{2}{15\pi}\hbar v_1 \left(\frac{\omega_p R}{c}\right)^5 \frac{\omega_p}{c}\int_0^\infty dt \int_0^\infty dy (y^2 + \beta_1^2 t^2)^{1/2} e^{-2\lambda y} \Delta''(y, \gamma t) f_1(y, t) \quad (7)$$

$$\frac{dQ}{dt} = -\frac{2}{15\pi}(\hbar v_1^2)\left(\frac{\omega_p R}{c}\right)^5 \int_0^\infty dt \int_0^\infty dy \, e^{-2\lambda y} \Delta''(y, \gamma t) f_2(y, t) \quad (8)$$

$$f_1(y, t) = \int_0^{2\pi} d\phi \cos\phi \frac{t_+}{(1 + t_+^2)}(y^2 + \beta_1^2 t_+^2/2)\left[\coth\frac{\alpha_1 t_+}{2} - \coth\frac{\alpha_1 w t}{2}\right] \quad (9)$$

$$f_2(y,t) = \int_0^{2\pi} d\phi \frac{t_+^2}{(1+t_+^2)}(y^2 + \beta_1^2 t_+^2/2)\left[\coth\frac{\alpha_1 t_+}{2} - \coth\frac{\alpha_1 wt}{2}\right] \tag{10}$$

$$\Delta_m''(y,t) = \operatorname{Im}\left[\frac{y - \sqrt{y^2 + t/(t+i)}}{y + \sqrt{y^2 + t/(t+i)}}\right] \tag{11}$$

The calculation results are shown in Figs. 2–3. Figure 2 compares the friction forces between a spherical gold particle and the gold surface as functions of temperature configurations (at $V = 1$ m/s) (a) and particles velocities (b), both at $z_0 = 10$ nm, $R = 3$ nm. Here and in Fig. 5 the base of the logarithm is 10.

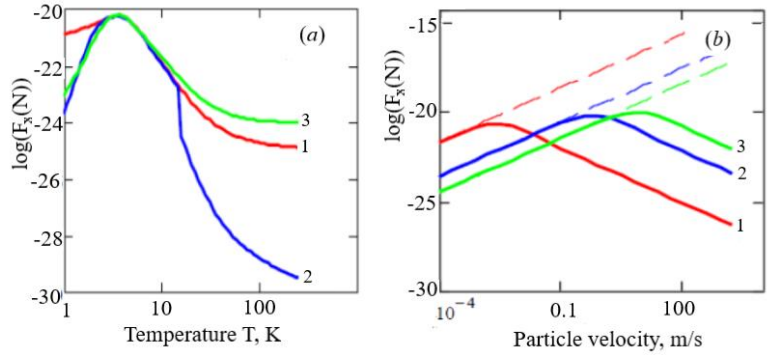

Fig. 2. Casimir-Lifshitz friction force between a gold particle and the gold surface ($z_0 = 10$ nm, $R = 3$ nm) as a function of temperature configuration – (a) (for $V = 1$ m/s) and velocity – (b). Different curves correspond to thermal configurations: (a) $T_1 = 3$ K, $T_2 = T$ (curve 1); $T_1 = T$, $T_2 = 3$ K (curve 2); $T_1 = T, T_2 = T$ (curve 3); (b) Curves 1–3 correspond to temperatures of 1, 3, 5 K for both the particle and the plate. The base of the logarithm is 10.

In general, the temperature difference causes a decrease in friction force, except for the case of curve 1 in Fig. 2a to the left of the maximum at $T_1 > T_2$. The presence of maxima in Fig. 2a,b is due to the already mentioned two factors: i) an increase in the contribution of low-frequency electromagnetic modes and ii) a change in the sign of the Doppler shifted frequency $\omega^+ = \omega + k_x V$, when $k_x < 0$. The effect of particle velocities demonstrates Fig. 2b. The dashed lines correspond to the linear-velocity approximation. We see that the lower the temperature, the narrower the range in which $F_x \propto V$. As a result, this simpler approximation can be used up to the velocity values corresponding to the maxima on the dependences $F_x(V)$. In summary, the peak value of the force increases slightly with increasing $T$ and $V$, but its position shifts to lower velocities with decreasing temperatures.

Dependences of $F_x(z_0)$ on the separation distance $z_0$ (not shown) are monotonically decreasing with increasing $z_0$. Typically, $F_x(a) \sim a^{-s}$, where the exponent index varies from $s = 1.3$ to $s = 2$, depending on the velocity and temperatures $T_{1,2}$. For room conditions ($T_{1,2} \sim 300$ K) and $V \sim 1$ m/s, where the dependence of $F_x$ on velocity $V$ is linear, it follows $s \approx 3$. This means that the slope of the dependence of friction force on distance decreases with decreasing temperature. This is consistent with [32–34], where we considered CL friction between gold plates at equilibrium conditions.

It is worth making a few remarks regarding possible experimental measurements of the Casimir-Lifshitz friction force. When measuring this force, a significant parameter is the ratio $\gamma_f = F_x/V$. According to Fig. 2, one obtains $\gamma_f \approx 10^{-17}$ kg/s as the maximum value at thermal equilibrium with $T = 1$ K, $V = 10^{-4}$ m/s (curve 1 in Fig. 2b). This is much closer to the experimental value $\gamma_f = 10^{-13} \div 10^{-14}$ kg/s in [45], than to the estimates [13, 25] in room conditions.

The key results demonstrating the anomalous radiative heat transfer are observed in Fig. 3a, corresponding to the case where the particle temperature is higher than that of the surface. One can see that when $V \to 0$, the particle is cooled and $\frac{dQ}{dt} < 0$. With increasing velocity $V$, the particle heats up the faster, the lower the temperature difference (curve 3). This is quite understandable, since the effect of heating due to photons with $k_x < 0$ is compensated in part by the "normally" directed heat flux from the plate. Figure 3b demonstrates the case of "normal" heat transfer, where the surface is hot. As the particle velocity increases, its heating rate also increases, but the order of the curves is opposite to Fig. 3a: $dQ/dt$ is larger at higher temperature difference (compare upper curve 3 in (a) with lower curve in (b)), and the absolute values of $dQ/dt$ in Fig. 3b are significantly lower.

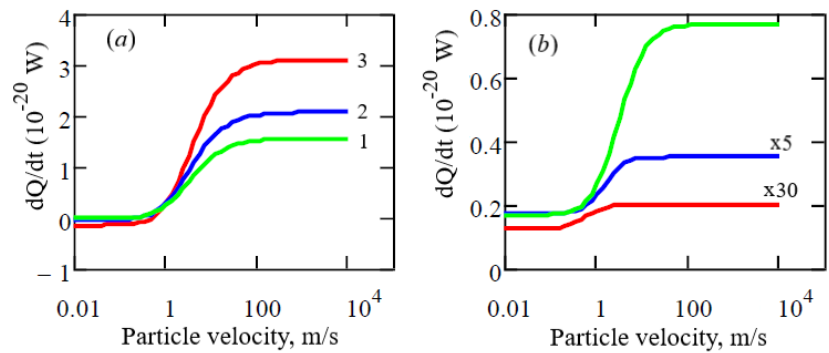

Fig. 3. Particle heating rate as a function of its velocity and temperature configuration ($z_0 = 10$ nm, $R = 3$ nm): (a) $T_1 = 4.2$, $T_2 = 1.5$ K (curve 1); $T_1 = 4.2$, $T_2 = 2$ K (curve 2); $T_1 = 4.2$, $T_2 = 3$ K (curve 3); (b) upper curve – $T_1 = 1.5$, $T_2 = 4.2$ K; middle curve – $T_1 = 2$, $T_2 = 4.2$ K; lower curve – $T_1 = 3$, $T_2 = 4.2$ K. The values for the two lower curves are increased by 5 and 30 times.

In addition, the maximum values of $dQ/dt$ on the plateaus in Fig. 3 significantly exceed the static RHT values (at $V \to 0$) by absolute values.

Figure 4 also manifests the presence of anomalous heating of particles until the temperature difference becomes too large. Shown is the $dQ/dt$ as a function of particle temperature $T_1$ at different velocities. The temperature of the plate is $T_2 = 4.2$ K. The curves from bottom to top correspond to $V = 0, 20, 50, 100$ m/s, respectively. The higher the speed, the greater the possible temperature difference. With a subsequent increase in $T_1 - T_2$, the $dQ/dt$ drops to its "normal" values corresponding to particle cooling (see lower curve at $V = 0$).

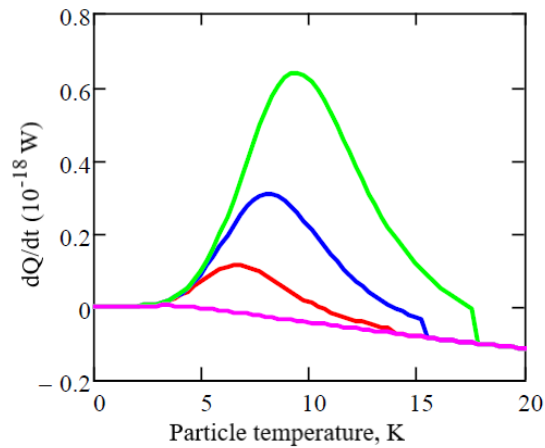

Fig. 4. Particle heating rate as a function of temperature $T_1$. The curves from bottom to top correspond to velocities $V = 0, 20, 50, 100$ m/s ($z_0 = 10$ nm, $R = 3$ nm, $T_2 = 4.2$ K).

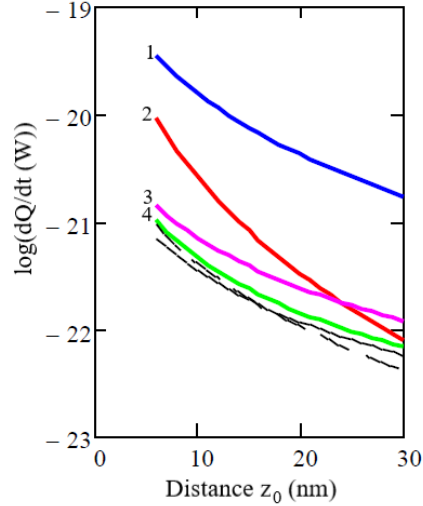

Fig. 5. Particle heating (cooling) rate as a function of separation distance $z_0$. Curves 1 and 3 correspond to velocities of 10 m/s, curves 2 and 4 – to velocities of 1 m/s, black solid and dashed curves – to $V = 0$. Temperature configurations: $T_1 = 4.2, T_2 = 2$ K (curves 1 and 2); $T_1 = 2, T_2 = 4.2$ K (curves 3 and 4); black solid curve ($\log(|dQ/dt|)$: $T_1 = 4.2, T_2 = 2$ K; black dashed curve: $T_1 = 2, T_2 = 4.2$ K. The base of the logarithm is 10.

The dependences of $dQ/dt$ on the distance $z_0$ are plotted in Fig. 5. Except for curve 2 ($V = 1$ m/s, $T_1 = 4.2, T_2 = 2$ K), all other curves have close monotonically decreasing slopes ($dQ/dt \propto z_0^{-s}$) with exponents $s \approx 1.7 \div 1.9$. Curve 2 has a larger slope $s \approx 2.95$. These features are explained by strong interplay of temperatures $T_1$, $T_2$, velocity $V$ and distance $z_0$. Curves 3, 4 and two lower curves in Fig. 5 correspond to the case of "normal" RHT direction. Curves 1 and 2 refer to the case of anomalous RHT, but the particle velocity in the second case is ten times less.

It should be noted that the absolute values of $dQ/dt$ in the dynamic mode turn out to be higher or of the same order of magnitude than in room conditions or at low temperatures in the static case. This can be seen when comparing the maximum values in Fig. 4 with the static results (lower line), and with the calculated values $|dQ/dt| = 1.04 \cdot 10^{-18}$ W and $4.2 \cdot 10^{-22}$ W, corresponding to configurations $T_1 = 300, T_2 = 4.2$ K and $T_1 = 4.2, T_2 = 300$ K, respectively (all for $z_0 = 10$ nm).

Thus, a temperature difference of several K between a moving metal particle and a metal surface causes a decrease in friction compared to the case of thermal equilibrium at temperatures of 1–20 K, although the friction force is approximately 3–4 orders of magnitude higher than at room conditions. Much more interesting is the fact that the direction of the RHT can change when the particle's velocity becomes large enough. This effect is explained by the action of counter-moving photons, which have a strong influence on both the friction force and radiative heat transfer, heating and the particle even if it is hotter than the surface. In particular, "anomalous Doppler effect" from counter-moving photons leads to the Cherenkov radiation of a high-speed particle moving parallel to a dielectric [27]. It is worthwhile noting that the effect of anomalous RHT is not a specific property of the Bloch–Grüneisen dependence $\nu(T)$. For metals, at low temperatures, when $\nu = const$ due to residual resistance, anomalous RHT also takes place [46]. Measurements of anomalous RHT seem to be very realistic, since the velocities of the probing body can be on the order of 1 m/s, and the RHT rate is higher or has the same order of magnitude than under close to room conditions. The low-temperature measurements of Casimir-Lifshitz friction force are also much more promising.

**Acknowledgments**

The author thanks C. Henkel for valuable comments and suggestions.

## Supplemental Material

### a) General relativistic formula for the rate of particle heating

In the problem of relativistic fluctuation-electromagnetic interaction of a small dipole particle moving with a constant velocity $V$ parallel to the surface of a thick plate with frequency-dependent dielectric permittivity $\varepsilon$ and magnetic permittivity $\mu$, one should discriminate the quantities relating to different inertial reference frames: the particle frame of rest, and that of the plate (laboratory system). The former one is co-moving with velocity $V$ in the $x-$ direction of the Cartesian coordinate system, associated with the plate. Then the rate of particle heating $dQ'/dt'$ in its frame of rest (i.e. local rate of heating) is given by Ref. [25]

$$\frac{dQ'}{dt'} = \frac{\hbar\gamma^3}{2\pi^2}\int_0^\infty d\omega \int d^2k \omega^+ \left[\alpha_e''(\gamma\omega^+)\mathrm{Im}\left(\frac{e^{-2q_0 z_0}}{q_0}R_e(\omega,\boldsymbol{k})\right) + (e\to m)\right]\left(\coth\frac{\hbar\omega}{2T_2} - \coth\frac{\hbar\gamma\omega^+}{2T_1}\right) \quad (S1)$$

where $\alpha_{e,m}''(\omega)$ are the imaginary parts of the particle electrical and magnetic polarizabilities, $\gamma = (1-V^2/c^2)^{-1/2} = (1-\beta^2)^{-1/2}$ is the Lorentz factor, $\omega^+ = \omega + k_x V$, $\boldsymbol{k} = (k_x, k_y)$, $q_0 = (k^2 - \omega^2/c^2)^{1/2}$, $q = (k^2 - \varepsilon\mu\,\omega^2/c^2)^{1/2}$, $R_e(\omega,\boldsymbol{k}) = A_1\Delta_e(\omega) + A_2\Delta_m(\omega)$, $R_m(\omega,\boldsymbol{k}) = A_1\Delta_m(\omega) + A_2\Delta_e(\omega)$,

$$A_1 = 2(k^2 - k_x^2\beta^2)(1-\omega^2/k^2c^2) + \frac{(\omega^+)^2}{c^2},$$

$$A_2 = 2k_y^2\beta^2(1-\omega^2/k^2c^2) + \frac{(\omega^+)^2}{c^2},$$

$$\Delta_e(\omega) = \frac{\varepsilon q_0 - q}{\varepsilon q_0 + q}, \qquad \Delta_m(\omega) = \frac{\mu q_0 - q}{\mu q_0 + q}.$$

In the limit $V/c \ll 1$ ($\gamma = 1$), taking the retardation into account, the quantity $dQ'/dt'$ coincides with the rate of particle heating $dQ/dt$ in the reference frame of the plate. In this case, formula (S1) reduces to (5). The expression for the CL friction force $F_x$ in the reference frame of the plate is obtained from (S1) when replacing the frequency multiplier $\omega^+$ by $(-k_x)$ in the integrand and $\gamma^3$ by $\gamma$ before the integral in (S1) (see Ref. [25])

### b) Structure of formula (5)

For $V = 0$, formula (5) coincides with the well-known results [12, 13, 25, 43]. For $V \neq 0$, its key feature is the presence of the frequency factor $\omega^+$ in the integrand. It is this factor that mathematically leads to the possibility of anomalous heating of the particle. To illustrate the appearance of $\omega^+$ in (5), we consider a simpler case of a nonretarded nonrelativistic interaction of a small particle with a fluctuating electric dipole moment. In this case, the initial expression for the particle heating rate $dQ/dt$ is given by [25]

$$dQ/dt = dQ^{(1)}/dt + dQ^{(2)}/dt = \langle \dot{\boldsymbol{d}}^{sp} \boldsymbol{E}^{ind} \rangle + \langle \dot{\boldsymbol{d}}^{ind} \boldsymbol{E}^{sp} \rangle, \tag{S2}$$

where indices "*sp*" and "*ind*" denote spontaneous and induced components of the fluctuating dipole moment of the particle and the electric field of the surface, the dots over $\boldsymbol{d}$ denote time derivatives, and the angular brackets denote complete quantum statistical averaging. When calculating the first term in (S2), the solution to the Poisson equation $\Delta\phi = 4\pi \mathrm{div} \boldsymbol{P}$ for the electric potential $\phi$ has to be find, where $\boldsymbol{P} = \delta(x - Vt)\delta(y)\delta(z - z_0)\boldsymbol{d}^{sp}(t)$ is the polarization created by fluctuating dipole moment $\boldsymbol{d}^{sp}(t)$ of a particle. The $\phi$ is expressed by the integral Fourier-transform

$$\phi(\boldsymbol{r}, z, t) = \frac{1}{(2\pi)^3} \int d\omega d^2 k \phi(\omega, \boldsymbol{k}; z) \exp(i(\boldsymbol{kr} - \omega t)), \tag{S3}$$

where $\boldsymbol{r} = (x, y), \boldsymbol{k} = (k_x, k_y)$. The $\boldsymbol{d}^{sp}(t)$ is expressed by

$$\boldsymbol{d}^{sp}(t) = \frac{1}{(2\pi)} \int d\omega \boldsymbol{d}(\omega) \exp(-i\omega t), \tag{S4}$$

The Poisson equation is solved under the standard boundary conditions $\phi(\boldsymbol{r}, +0, t) = \phi(\boldsymbol{r}, -0, t), \partial_z(\boldsymbol{r}, z, t)_{z=+0} = \varepsilon \partial_z(\boldsymbol{r}, z, t)_{z=-0}$, where $\varepsilon$ is the dielectric permittivity of the plate. For the Fourier- component of the induced potential created by the moving particle, it follows [25]

$$\phi^{ind}(\omega, \boldsymbol{k}; z) = \frac{2\pi}{k} \Delta(\omega) \exp(-k(z + z_0))\left[i \boldsymbol{k} \boldsymbol{d}^{sp}(\omega - k_x V) + k d_z^{sp}(\omega - k_x V)\right], \tag{S5}$$

where $\Delta(\omega) = (\varepsilon(\omega) - 1)/(\varepsilon(\omega) + 1)$. Using (S5) and the relationship $\boldsymbol{E}^{ind} = -\nabla\phi^{ind}$, the induced electric field at the particle location point $(Vt, 0, z_0)$ is given by

$$\boldsymbol{E}^{ind} = \frac{1}{(2\pi)^3} \int d\omega d^2 k \phi^{ind}(\omega, \boldsymbol{k}; z) \exp(-i(\omega - k_x V)). \tag{S6}$$

Having substituted (S4) and (S6) into the first term of (S2) and taking the correlator of the particle dipole moments into account

$$\langle d_i^{sp}(\omega) d_k^{sp}(\omega') \rangle = 2\pi \delta_{ik} \, \hbar \delta(\omega + \omega') \alpha''(\omega) \coth\frac{\hbar\omega}{2T_1}, \tag{S7}$$

where $i, k = x, y, z$ and $\alpha''(\omega)$ is the imaginary part of the particle polarizability, we obtain

$$\frac{dQ^{(1)}}{dt} = -\frac{\hbar}{\pi^2}\int_0^\infty d\omega \int_{-\infty}^{+\infty} dk_x \int_{-\infty}^{+\infty} dk_y k\omega^+ e^{-2kz_0}\alpha''(\omega^+)\text{Im}(\omega)\coth\frac{\hbar\omega^+}{2T_1}. \tag{S8}$$

where $\omega^+ = \omega + k_x V$. When obtaining (S8), the analytical properties of the functions $\alpha(\omega)$ and $\Delta(\omega)$ are used (evenness of their real parts and oddness of imaginary parts). When calculating the term $dQ^{(2)}/dt$ in (S2), the linear integral relation between $\boldsymbol{E}^{sp}$ and $\boldsymbol{d}^{ind}$ is used, yielding

$$\boldsymbol{d}^{ind}(t) = \frac{1}{(2\pi)^3}\int d\omega \alpha(\omega - k_x V)\boldsymbol{E}^{sp}(\omega, \boldsymbol{k}; z_0)\exp(-(\omega - k_x V)t). \tag{S9}$$

The correlator of the electric fields of the plate arising in this case is worked out using the fluctuation-dissipation relation [25]

$$\langle \boldsymbol{E}^{sp}(\omega, \boldsymbol{k}; z_0)\boldsymbol{E}^{sp}(\omega', \boldsymbol{k}'; z_0)\rangle = 2(2\pi)^4 \hbar k e^{-2kz_0}\Delta''(\omega)\delta(\omega + \omega')\delta(\boldsymbol{k} + \boldsymbol{k}')\coth\frac{\hbar\omega}{2T_2}. \tag{S10}$$

Substituting (S9) and (S10) into the second term of (S2) yields

$$\frac{dQ^{(2)}}{dt} = \frac{\hbar}{\pi^2}\int_0^\infty d\omega \int_{-\infty}^{+\infty} dk_x \int_{-\infty}^{+\infty} dk_y k\omega^+ e^{-2kz_0}\alpha''(\omega^+)\text{Im}(\omega)\coth\frac{\hbar\omega}{2T_2}. \tag{S11}$$

Summing up (S8) and (S11), yields

$$\frac{dQ}{dt} = \frac{\hbar}{\pi^2}\int_0^\infty d\omega \int_{-\infty}^{+\infty} dk_x \int_{-\infty}^{+\infty} dk_y k\omega^+ e^{-2kz_0}\alpha''(\omega^+)\text{Im}(\omega)\left(\coth\frac{\hbar\omega}{2T_2} - \coth\frac{\hbar\omega^+}{2T_1}\right). \tag{S12}$$

Formula (S12) coincides with (5) when using the non-retarded limit ($c \to \infty, q_0 \to k$) in (5). From the relativistic solution of this problem it follows $\Delta(\omega) \to (\varepsilon q_0 - q)/(\varepsilon q_0 + q)$. Similarly, if the heating of the particle due to the magnetic polarizatio is taken into account, one obtains $dQ/dt = \langle \dot{\boldsymbol{m}}^{sp}\boldsymbol{B}^{ind}\rangle + \langle \dot{\boldsymbol{m}}^{ind}\boldsymbol{B}^{sp}\rangle$, where $\boldsymbol{m}^{sp,ind}$ and $\boldsymbol{B}^{sp,ind}$ are the spontaneous and induced components of the fluctuation magnetic moment of the particle and the magnetic field of the plate. The corresponding calculations lead to formula (S11) with the electric polarizability replaced by the magnetic one and $\Delta(\omega) \to (\mu - 1)/(\mu + 1)$. Upon relativistic consideration, respectively, $\Delta(\omega) \to (\mu q_0 - q)/(\mu q_0 + q)$, and we obtain Eq. (3) for $\Delta_m$ with $\mu = 1$.

Thus, the appearance of "shifted" frequency $\omega^+$ in the formulas for the heating rate $dQ/dt$ of a particle is mathematically due to the presence of derivatives of the dipole moment in (S2), which must be taken before substituting the instantaneous coordinates of the particle $(Vt, 0, z_0)$, with subsequent application of analytical properties of polarizability $\alpha(\omega)$ and Fresnel's reflection coefficient $\Delta(\omega)$.